\documentclass[twocolumn,conference,letterpaper]{IEEEtran}
\usepackage[left=0.75in,right=0.75in,top=1in,bottom=0.75in]{geometry}

\usepackage{cite}
\usepackage{amssymb,amsmath,latexsym,amsfonts,amsthm,mathtools}

\usepackage{graphicx}
\usepackage{float,subcaption}
\usepackage{textcomp}
\usepackage{xcolor}
\definecolor{darkpastelpurple}{rgb}{0.59, 0.44, 0.84}
\usepackage{hyperref}
\hypersetup{colorlinks, breaklinks, citecolor=blue, linkcolor=blue, urlcolor=blue}
\usepackage[nameinlink]{cleveref}
\Crefformat{figure}{#2Fig.~#1#3}
\Crefmultiformat{figure}{Figs.~#2#1#3}{ and~#2#1#3}{, #2#1#3}{ and~#2#1#3}

\crefname{assumption}{assumption}{assumptions}  
\Crefname{assumption}{Assumption}{Assumptions} 
\usepackage{tikz}
\usetikzlibrary{automata, shapes, arrows, calc, arrows.meta, fit, positioning}

\theoremstyle{plain}
\newtheorem{theorem}{Theorem}
\newtheorem{lemma}{Lemma}
\newtheorem{corollary}{Corollary}

\newtheorem*{problem*}{Problem}
\theoremstyle{remark}
\newtheorem{remark}{Remark}

\newtheorem{definition}{Definition}
\theoremstyle{definition}

\usepackage{mathrsfs}

\usepackage{newtxmath}
\usepackage{booktabs}
\usepackage{duckuments}

\IEEEoverridecommandlockouts

\begin{document}
	\title{Trajectory Encryption Cooperative Salvo Guidance}
	\author{Lohitvel Gopikannan, Shashi Ranjan Kumar,~\IEEEmembership{Senior Member, IEEE}, and Abhinav Sinha,~\IEEEmembership{Senior Member,~IEEE}
		\thanks{L. Gopikannan and S. R. Kumar are with the Intelligent Systems \& Control (ISaC) Lab, Department of Aerospace Engineering, Indian Institute of Technology Bombay, Mumbai 400076, India. (e-mails: 24m0023@iitb.ac.in, srk@aero.iitb.ac.in). A. Sinha is with the Guidance, Autonomy, Learning, and Control for Intelligent Systems (GALACxIS) Lab, Department of Aerospace Engineering and Engineering Mechanics, University of Cincinnati, OH 45221, USA. (e-mail: abhinav.sinha@uc.edu).}
	}

	\maketitle
	\thispagestyle{empty}
	
	\begin{abstract}
		This paper introduces the concept of trajectory encryption in cooperative simultaneous target interception, wherein heterogeneity in guidance principles across a team of unmanned autonomous systems is leveraged as a strategic design feature. By employing a mix of heterogeneous time-to-go formulations leading to a cooperative guidance strategy, the swarm of vehicles is able to generate diverse trajectory families. This diversity expands the feasible solution space for simultaneous target interception, enhances robustness under disturbances, and enables flexible time-to-go adjustments without predictable detouring. From an adversarial perspective, heterogeneity obscures the collective interception intent by preventing straightforward prediction of swarm dynamics, effectively acting as an encryption layer in the trajectory domain.   Simulations demonstrate that the swarm of heterogeneous vehicles is able to intercept a moving target simultaneously from a diverse set of initial engagement configurations.
	\end{abstract}
	
	
	\section{Introduction}\label{sec:introduction}
	Cooperative intercept missions, once limited to large-scale interceptor systems, are also being realized using agile teams of small drones. This shift introduces both new opportunities and challenges. Small drones offer modular flexibility, scalability, and low-cost deployment. However, the requirement of cooperative simultaneous target interception demands precise coordination across vehicles with varying dynamics, constraints, and operating requirements. This demand for precise coordination marks a departure from traditional non-cooperative salvo tactics. 
	
	Early cooperative guidance strategies often focused on the simultaneous interception of stationary targets. For instance, the authors in \cite{sp11} formulated a distributed cooperative guidance law leveraging biased proportional navigation (PN), while the work in \cite{sp12} re-conceptualized impact-time control as a range-tracking problem. To mitigate communication burdens, the work in \cite{sp7} devised finite-time distributed protocols that exclusively utilized local time-to-go estimates. Similarly, that in \cite{sp15} formulated a receding-horizon cooperative framework wherein interceptors exchanged information solely with their neighbors to solve localized, finite-horizon optimization problems. Concurrently, alternative approaches in \cite{sp16,sp17} circumvented explicit time-to-go estimation by employing a two-stage framework that first established a decentralized consensus on range and heading, subsequently followed by PN guidance. Although these strategies successfully achieved consensus over undirected network topologies, their convergence was asymptotic, which can compromise performance in time-critical, short-duration engagements.

	A leader-follower architecture, presented in \cite{sp1}, leveraged PN-based time-to-go estimation and super-twisting sliding mode control to guarantee convergence within a finite time for the simultaneous interception of stationary targets, even in the presence of large heading errors. A critical vulnerability of any leader-follower framework is its sensitivity to the leader's failure, which could jeopardize the mission's success. \cite{hp6} proposed a cooperative salvo guidance strategy based on sliding-mode fixed finite-time consensus, transforming the problem into a consensus framework over a cycle digraph with heterogeneous gains. To explicitly account for acceleration constraints, a nonlinear cooperative guidance law was developed in \cite{hp13} which maintained stability even for large heading errors and demonstrated the applicability across both directed and undirected network topologies, though its consensus convergence remained asymptotic in nature. A leader–follower cooperative salvo guidance framework was proposed in \cite{sp26}, employing predefined-time consensus on time-to-go to achieve simultaneous interception under nonlinear kinematics and autopilot dynamics. Predefined-time consensus is particularly suited for time-critical engagements, as it ensures convergence within a user-specified duration. Studies in both \cite{sp1} and \cite{hp6} extended their guidance strategies to moving targets using the predicted interception point method. However, since time-to-go is inherently formulated for stationary targets, applying it to moving targets requires approximating the target’s future position, which can introduce errors and reduce interception efficacy

	For interception of moving targets, studies in \cite{hp11,hp12} developed finite-time consensus–based salvo guidance over undirected and directed graphs, but its dependence on radial acceleration for time-to-go synchronization hinders practical implementation. The work in \cite{hp10} proposed cooperative salvo guidance using asymptotic consensus over undirected graphs against unknown target maneuvers. Time-constrained interception under deviated pursuit has been studied in \cite{hp4,hp7}, where an exact analytical time-to-go against constant-velocity targets enabled precise interception via deviation angle modulation. Building on this, the framework has been extended in \cite{hp8} to simultaneous interception of moving targets. By leveraging weighted consensus on time-to-go with suitably designed edge weights, including admissible negative values, the approach enables flexible and precise time-constrained interception beyond the convex hull of initial estimates. In \cite{hp5}, true proportional navigation guidance was introduced to overcome the limitation of deviated pursuit against stationary targets, extending its use to a wider target class. A cooperative scheme was then proposed to enforce time-to-go consensus for simultaneous interception, with the interception time either pre-set or determined cooperatively. A leaderless cooperative guidance strategy based on deviated pursuit and a true proportional navigation guidance framework for simultaneous interception of fast, non-accelerating targets is proposed in \cite{hp9}.
	
	Note that traditional approaches presented in the aforementioned works utilize homogeneous guidance principles, which may simplify the guidance design but impose structural limitations, yielding predictable trajectories, high coordination burdens to satisfy certain common impact time requirements, and vulnerability to adversarial inference. To overcome these drawbacks, we propose a cooperative strategy based on heterogeneity in guidance principles, thereby enabling trajectory diversity, maneuvering flexibility, and robustness to the target's motion. Moreover, such heterogeneity obscures collective intent from an adversarial perspective, effectively acting as a form of trajectory encryption that enhances both mission effectiveness and operational security, while guaranteeing cooperative simultaneous interception.
	
	In this paper, we consider a cooperative simultaneous interception of a fixed-wing drone using fixed-wing and multirotor vehicles. However, any set of heterogeneous vehicles can be accommodated using the proposed strategy. The fixed-wing vehicles in the swarm have the ability to modulate their bearing angles to a constant value for time-constrained target interception, thus utilizing deviated pursuit. The multirotors, on the other hand, are endowed with the capability to modulate their thrust or propulsion, thus utilizing both radial and tangential components of their accelerations following true proportional navigation. Although different classes of vehicles have different time-to-go formulations, we show that consensus in their heterogeneous time-to-go values is established within a predefined time via the proposed method, regardless of the initial engagement geometry. 
	
	Moreover, the proposed strategy is not limited to assigning fixed guidance laws to individual vehicles. It is flexible enough to allow morphing of the guidance commands during the engagement if the vehicles can issue commands following different guidance principles to dynamically switch or blend between them in real time. For instance, a vehicle may initially employ pure proportional navigation to be energy-efficient, but may transition to aggressive nonlinear pursuit guidance to fine-tune timing and counter evasive maneuvers in the endgame. 
	
	Therefore, the proposed strategy renders the choice of guidance law as a tunable degree of freedom by augmenting trajectory diversity beyond static heterogeneity to ensure adaptability in contested environments. From a cooperative standpoint, this morphing capability provides an additional layer of resilience. Even if adversaries infer a vehicle’s current behavior, the underlying principle can change mid-flight, complicating prediction and interception. Consequently, the system achieves both trajectory encryption through structural diversity and maneuver flexibility through command morphing, enabling robust and deceptive time-critical coordination in contested environments.
	
	\section{Preliminaries and Problem Formulation}
\begin{figure}[h!]
		\centering
		\includegraphics[width=0.84\linewidth]{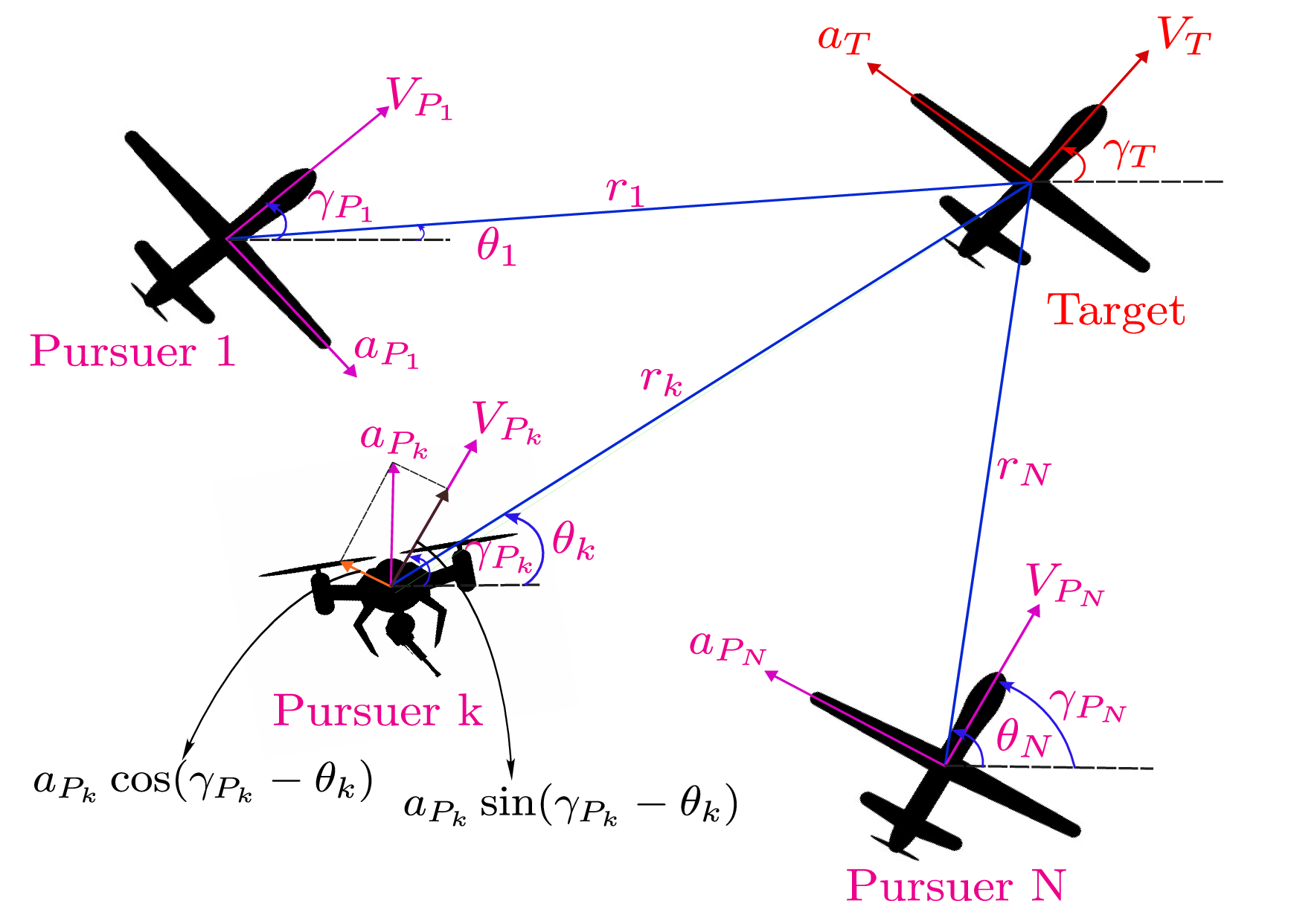}
		\caption{Cooperative heterogeneous multi-agent engagement.}
		\label{fig:enggeo}
	\end{figure}

	A cooperative engagement scenario with $N$ pursuers against a non-maneuvering constant velocity target, as shown in \Cref{fig:enggeo}, is considered, wherein heterogeneity arises from the presence of different classes of vehicles that can adopt distinct guidance strategies. In this paper, we consider $M$ fixed-wing pursuers employing deviated pursuit guidance (DPG), while the remaining $N-M$ pursuers are multirotors, which follow true proportional navigation guidance (TPNG). The essential difference arises from their actuation frameworks. A DPG pursuer can generate control solely through the lateral acceleration channel, whereas a TPN pursuer exploits both radial and lateral acceleration channels to achieve guidance. However, any other guidance principles may also be adopted.
	
	Let $V_T$ and $\gamma_T$ denote the velocity and flight path angle of the target, respectively. For the $i^{\text{th}}$ pursuer, the velocity is represented by $V_{P_i}$ and the flight path angle by $\gamma_{P_i}$. The line-of-sight (LOS) angle between the $i^{\text{th}}$ pursuer and the target is denoted by $\theta_i$, whereas the relative distance between the $i$\textsuperscript{th} agent and the target is $r_i$. Their evolution is given by
	\begin{subequations}\label{eq:enggeo}
		\begin{align}
			\dot{r}_i &= V_T \cos(\gamma_T - \theta_i) - V_{P_i} \cos\delta_i=V_{r_i} , \\
			r_i\dot{\theta}_i&= V_T \sin(\gamma_T - \theta_i) - V_{P_i} \sin\delta_i=V_{\theta_i},
		\end{align}
	\end{subequations}
	where $V_{r_i}$ and $V_{\theta_i}$ represents relative velocity components along and across the LOS. For the $i$\textsuperscript{th} DPG pursuer, the control input is its lateral acceleration, that is,
	\begin{equation}
		\dot \gamma_{P_i}=\dfrac{a_{P_i}}{V_{P_i}};~\forall~i \in \{1,\ldots,M\},
	\end{equation}
	whereas the TPNG pursuers have radial and tangential acceleration as control inputs, given by
	\begin{equation}
		\dot{\gamma}_{P_j} = \frac{a_{P_j} \cos(\gamma_{P_j} - \theta_j)}{V_{M_j}}, ~
		\dot{V}_{P_j} = a_{P_j} \sin(\gamma_{P_j} - \theta_j),
	\end{equation}
	$\forall~j\in\{M+1,\ldots,N\}$. In this paper, the target and the pursuers are treated as ideal point-mass vehicles. With a sufficiently fast autopilot, the pursuers' dynamics may be ignored when designing the guidance strategy.
	\begin{problem*}
		Design a distributed cooperative guidance strategy for heterogeneous pursuers connected over a directed communication topology to ensure synchronized interception of a non-maneuvering constant velocity target.
	\end{problem*}
	The vehicles' (agents) interaction can be modeled via graphs. Since bidirectional communication is often impractical, directed topologies offer a more efficient and feasible alternative in real-world salvo engagements. Therefore, let the communication topology among the pursuers be modeled by a strongly connected directed graph $\mathcal{G}=\{\mathcal{V},\mathcal{E}\}$, where the node set is defined as $\mathcal{V}=\{v_1, v_2, \ldots, v_N\}$, representing the $N$ pursuers. The directed edge set $\mathcal{E}\subseteq \mathcal{V}\times \mathcal{V}$ characterizes the information flow, where an edge $e_{ij}\in \mathcal{E}$ denotes a directed communication link from agent $v_i$ to agent $v_j$. For each node $v_i\in \mathcal{V}$, the in-neighbor set is given by $\mathcal{N}_i=\{v_j\in\mathcal{V}\;|\;e_{ji}\in\mathcal{E}\}$, which represents all agents from which $v_i$ receives information. The communication structure is encoded in the adjacency matrix $\mathcal{A}=[a_{ij}]\in\mathbb{R}^{N\times N}$, with 
	\[
	a_{ij}=\begin{cases}
		1,& \text{if } e_{ji}\in \mathcal{E},\\
		0,& \text{otherwise},
	\end{cases},~\text{and } a_{ii}=0.
	\]
	The in-degree matrix is defined as $\mathcal{D}=\text{diag}\{d_1,d_2,\ldots,d_N\}$, where each diagonal entry $d_i=\sum_{j=1}^{N}a_{ij}$ represents the number of incoming edges to node $v_i$. The associated graph Laplacian is then given by $\mathcal{L}=\mathcal{D}-\mathcal{A}$, and the Kronecker product $\otimes$ is used to construct higher-dimensional system dynamics from scalar graph structures.
	\begin{definition}[\cite{hp2}]\label{def:assignabletime}
		Consider the nonlinear system
		\begin{equation}\label{system}
			\dot{x}(t) = f(t, x(t)), \quad t \in \mathbb{R}_+, \quad x(0) = x_0,
		\end{equation}
		where $x \in \mathbb{R}^n$ denotes the state vector and $f : \mathbb{R}_+ \times \mathbb{R}^n \to \mathbb{R}^n$ is a nonlinear mapping, uniformly locally bounded in time.  
		The equilibrium at the origin of \eqref{system} is said to be {globally prescribed-time stable} if it is globally uniformly finite-time stable, and the convergence time $T$ can be arbitrarily assigned as a finite user-specified constant. More precisely, for any $0 < T_p \leq T_{\max} < \infty$, there exists a prescribed settling time $T$ satisfying $
		T_p \leq T \leq T_{\max}, \quad \forall \, x_0 \in \mathbb{R}^n,$
		where $T_p$ represents the minimum physically achievable convergence time.
	\end{definition}
	Before proceeding further, we define a time-varying scaling function \cite{hp1} as follows,
	\begin{align}\label{timevarying}
		h(t) =
		\begin{cases}
			\displaystyle \frac{t_e}{\pi} \sin\!\left(\frac{\pi}{t_e} t \right) + t - t_e, & 0 \leq t < t_e, \\[6pt]
			1, & t \geq t_e,
		\end{cases}
	\end{align}
	where $t_e > 0$ denotes a user-specified convergence time and $h(t)<0$ for $t \in [0,t_e)$. The corresponding time derivative is 
	\begin{align}\label{timevaryingdot}
		\dot{h}(t) =
		\begin{cases}
			\displaystyle \cos\!\left(\frac{\pi}{t_e} t \right) + 1, & 0 \leq t < t_e, \\[6pt]
			0, & t \geq t_e.
		\end{cases}
	\end{align}	
	\begin{lemma}[\cite{hp2}]\label{lem:ft}
		Let $V(x(t), t) : \mathcal{D} \times \mathbb{R}_+ \to \mathbb{R}$ 
		be a continuously differentiable function, where $\mathcal{D} \subset \mathbb{R}^n$ is a domain containing the origin.  Assume there exists a constant $b > 0$ such that
		$
		V(0, t) = 0, 
		\quad 
		V(x(t), t) > 0, \;\; \forall \, x(t) \in \mathcal{D} \setminus \{0\},
		$
		and
		$
		\dot{V}(x(t), t) \;\leq\; -\,bV(x(t), t) - 2 \frac{\dot{\mu}(t)}{\mu(t)} V(x(t), t), 
		\quad \forall \, x(t) \in \mathcal{D},
		$
		for all $t \in [t_0, \infty)$.  
		Then, the origin of system~\eqref{system} is \emph{prescribed-time stable}, with the convergence time $T$ determined by the user-defined function $\mu(t)$ satisfying $\dfrac{\dot{\mu}(t)}{\mu(t)} > 0$.  If $\mathcal{D} = \mathbb{R}^n$, the origin of the system is globally prescribed-time stable under the same condition.  
		Furthermore, for $t \in [t_0, t_1)$, one has
		$
		V(t) \;\leq\; \mu^{-2}(t)\, \exp\!\big(-b(t - t_0)\big)\, \bigl(\mu(t_0)\bigr)^2 V(t_0),
		$
		while for $t \in [t_1, \infty)$ it holds that $V(t) \equiv 0$.
	\end{lemma}
	
	\begin{lemma}[\cite{hp3}]
		Consider a multi-agent network with a fixed communication topology $\mathcal{G}$ represented by a strongly connected digraph.  
		Let $\mathcal{L}$ denote the Laplacian matrix associated with $\mathcal{G}$, and let $\hat{\mathcal{L}}$ represent the Laplacian matrix of the corresponding undirected mirror graph $\hat{\mathcal{G}}$.  
		Then, the inequality $x^\top \mathcal{L} x \;\geq\; \lambda_2(\hat{\mathcal{L}})\,\|x\|^2$ holds, where $\lambda_2(\hat{\mathcal{L}})$ is the Fiedler eigenvalue of $\hat{\mathcal{G}}$, characterizing the algebraic connectivity of the mirror graph.
	\end{lemma}

	\section{Trajectory Encryption Guidance Strategy}
	In this section, we first introduce certain definitions related to interception time, followed by the derivation of the distributed cooperative trajectory encryption guidance strategy.
	\begin{definition}\label{def:inttime}
		Interception time for the $i$\textsuperscript{th} pursuer, $t_{f_i}$, is defined as the time taken by it to intercept the target, expressed as $t_{f_i} = t_{{\rm go}_i} + t_{\rm el}$, where $t_{{\rm go}_i}$ denotes its time-to-go, and $t_{\rm el}$ represents the elapsed time.
	\end{definition}
	The proposed strategy leverages this heterogeneity to develop a cooperative guidance strategy. Each pursuer may generate control inputs according to its own guidance principle, and as will be shown later, the proposed framework is flexible enough to allow morphing of guidance commands mid-engagement. From a security standpoint, this dynamic and heterogeneous use of guidance laws provides an implicit layer of trajectory encryption, allowing a tactical advantage to the interceptor swarm.
	
	Depending on the engagement configuration, each pursuer has a different estimate of its interception time, that is, $t_{f_i}\neq t_{f_j}~\forall i,j\in\{1,2,\ldots, N\}$ and $i\neq j$ in general. However, for a simultaneous target interception, each pursuer must agree on a common interception time as engagement proceeds (without manual intervention or pre-programming). To this end, control over their time-to-go is necessary.
	\begin{definition}\label{def:tgo_error}
		Time-to-go error for the $k$\textsuperscript{th} pursuer is given by
		\begin{equation}
			e_k(t) = t_{{\rm go}_k} - t_{{\rm go}_k}^d, \label{tgoerror}
		\end{equation}
		where $t_{{\rm go}_k}$ is the time-to-go of the $k$\textsuperscript{th} pursuer and $t_{{\rm go}_k}^d$ is its desired time-to-go (not prescribed explicitly). 
	\end{definition}
	Note that because the pursuers are heterogeneous, they have non-identical formulations for their time-to-go, and hence, could generate diverse and evolving trajectory patterns. Despite that, the pursuers must synchronize their time-to-go values through a distributed consensus mechanism under such heterogeneity.
	
	On differentiating \eqref{tgoerror} with respect to time, we get
	\begin{equation}
		\dot e_k = \dot t_{{\rm go}_k}+1, \label{tgoerrordot}
	\end{equation}
	Assume the time-to-go dynamics for the $k$\textsuperscript{th} pursuer can be written in affine form
	\begin{equation}\label{eq:tgo_affine}
		\dot t_{{\rm go}_k} = F_k(\mathbf{x}_k,t) + B_k(\mathbf{x}_k,t)\,a_{P_k}
	\end{equation}
	where $F_k$ and $B_k$ are agent-dependent scalar functions reflecting the dynamics following a specific guidance principle, and $\mathbf{x}_k$ is the set of relevant states constructed based on engagement variables. Substituting \eqref{eq:tgo_affine} in the \eqref{tgoerrordot} leads to the error dynamics
	\begin{equation}\label{eq:edot_affine}
		\dot e_k = 1 + F_k(\mathbf{x}_k,t) + B_k(\mathbf{x}_k,t)\,a_{P_k}.
	\end{equation}
	\begin{theorem}
		Consider a cooperative target engagement scenario involving heterogeneous pursuers, wherein different pursuers exchange information over a directed graph $\mathcal{G}$, and adopt different guidance principles. The guidance command for the $k$\textsuperscript{th} pursuer that leads to consensus in time-to-go within a predefined-time $t_e$, regardless of initial engagement configuration, is given by
		\begin{align}
			a_{P_k} = \dfrac{{-1}-F_k+\left(-\alpha + \beta \dfrac{\dot{h}(t)}{h(t)}\right)\displaystyle \sum _{j=1}^N[\mathcal{L}]_{kj} e_j}{B_k},\label{eq:aPk}
		\end{align}
		for some  $\alpha>0,\beta\geq{1}/({   \lambda_2(\mathcal{\hat{\mathcal{L}}})})$ and $h(t),\dot{h}(t)$ are defined in \eqref{timevarying}-\eqref{timevaryingdot}. Consequently, the heterogeneous pursuers intercept the moving target simultaneously at a time $t_f$ decided via implicit coordination.
	\end{theorem}
	\begin{proof}
		Let us consider $\mathbf{e}=    \begin{bmatrix}
			e_1 & e_2 & \cdots & e_N
		\end{bmatrix}^\top$ and $\dot{\mathbf{e}}=    \begin{bmatrix}
			\dot e_1 & \dot e_2 & \cdots & \dot e_N
		\end{bmatrix}^\top.$ Consider the time-to-go dynamics for the $k$\textsuperscript{th} pursuer \eqref{eq:tgo_affine}. Upon substituting the cooperative guidance command \eqref{eq:aPk} into \eqref{eq:tgo_affine}, one may obtain
		\begin{equation}\label{eq:generrdyn}
			\dot{\mathbf{e}} = \left(-\alpha + \beta \dfrac{\dot{h}(t)}{h(t)}\right)\displaystyle \mathcal{L}\mathbf{e}.
		\end{equation}
		Let us choose a Lyapunov function candidate $V=\dfrac{\mathbf{e}^\top\mathbf{e}}{2}$, whose time derivative is
		\begin{align}
			\dot V&= \mathbf{e}^\top\dot{\mathbf{e}}=\mathbf{e}^\top\left(-\alpha + \beta \frac{\dot{h}(t)}{h(t)}\right)\mathcal{L}\mathbf{e}\nonumber\\
			&= -\mathbf{e}^\top\alpha \mathcal{L}\mathbf{e}+\mathbf{e}^\top\beta \dfrac{\dot h(t)}{h(t)}\mathcal{L}\mathbf{e}  \notag\\
			&\leq -\alpha\lambda_2(\hat{\mathcal{L}})\|\mathbf{e}\|^2+\beta\frac{\dot h(t)}{h(t)}\lambda_2(\hat{\mathcal{L}})\|\mathbf{e}\|^2 .\label{eq:Vdotstep1}
		\end{align}
		Since $\|\mathbf{e}\|^2 = 2V$, \eqref{eq:Vdotstep1} can be further simplified to
		\begin{align}
			\dot V&\leq -2\alpha\lambda_2(\hat{\mathcal{L}}) V+2\beta \frac{\dot h(t)}{h(t)} \lambda_2(\mathcal{\hat L})V.\label{eq:Vdotstep2}
		\end{align}
		Now, consider $\hat \alpha =2\alpha\lambda_2(\hat{\mathcal{L}}) $ and $g(t)=\dfrac{1}{h(t)}$. Then, we get $\dfrac{\dot{h}(t)}{h(t)}=-\dfrac{\dot{g}(t)}{g(t)}>0$. By design $\alpha >0$ and $\beta\geq\dfrac{1}{\lambda_2(\mathcal{\hat L})}$, and hence on substituting these into \eqref{eq:Vdotstep2}, one may obtain
		\begin{align}\label{v_inequality}
			\dot V&\leq- \hat \alpha V-{2} \frac{\dot g(t)}{g(t)} V.
		\end{align}
		From \Cref{lem:ft}, we can write
		\begin{align}
			g^2(t) V(t) &\leq \exp\left( -\hat{\alpha}t \right) g^2(0) V(0), \nonumber\\
			=& \frac{1}{t_e^2}\exp\left( -\hat{\alpha}t \right)V(0)  ,
		\end{align}
		which is equivalent to writing
		\begin{align}\label{v_final}
			V(t)\leq \frac{1}{g^2(t)\;t_e^2}\exp\left( -\hat{\alpha}t \right)V(0). \end{align}
		From \eqref{v_inequality}, we can observe that the system is globally asymptotically stable for $t\in(0,\infty)$ and from \eqref{v_final} it can be inferred that  $V(t) \equiv 0$, which implies that $e(t) \equiv 0$ as $t \rightarrow t_e$. Therefore, from \Cref{lem:ft} we can conclude that the time-to-go error dynamics \eqref{eq:edot_affine} is globally prescribed time stable under the proposed command $a_{P_k}$. This essentially means that a consensus is established in the pursuers' time-to-go values within a time $t_e$ regardless of the initial engagement configuration. Thereafter, since $e_k$ is decreasing to zero, so is $t_{go_k}$, leading to a simultaneous interception.
	\end{proof}
	\begin{remark}
		The proposed prescribed-time consensus protocol inspired by \cite{hp1} can be seen in the error dynamics \eqref{eq:generrdyn}. The command \eqref{eq:aPk} can be alternatively expressed as
		\begin{align}
			a_{P_k} = -\dfrac{{1}+F_k}{B_k} + \sum_{j=1}^{N}w_{kj}e_j,
		\end{align}
		where $w_{kj}$ (time-varying) is agent-specific, conditioned on the heterogeneity and the network topology, which reflects the heterogeneous contributions from neighbors. The proposed framework is flexible enough to permit guidance morphing, wherein a pursuer dynamically switches or blends between multiple guidance laws during engagement. This cooperative framework \eqref{eq:aPk} functions as a strategic overlay, ensuring simultaneous interception in a way that allows any underlying guidance principle to be used interchangeably. The proposed approach provides a tactical advantage by obscuring the collective interception objective and complicating adversarial prediction, while simultaneously ensuring that mission success remains independent of the specific guidance laws employed or the diversity of the participating vehicles.
	\end{remark}
	From an adversary's perspective, the interceptor swarm produces a state trajectory $\mathbf{X}(\mathbf{E},\mathscr{G})\coloneqq \left\{\mathbf{x}_k(\cdot, a_{P_k})\right\}_{k=1}^N$, where $\mathbf{E}$ is the set containing the relevant engagement variables for the entire swarm, and $\mathscr{G}$ is the swarm guidance strategy composed of individual guidance commands. The measurement model for an adversary is typically expressed as
	\begin{align}
		\mathcal{O} = \mathcal{C}\left(\mathbf{X}(\mathbf{E},\mathscr{G})\right)+\mathbf{w},
	\end{align}
	where $\mathcal{C}$ is an observation operator (e.g., positions, bearing angles, LOS, etc.) and $\mathbf{w}$ is measurement noise. The adversary’s inference target could be a discrete or continuous variable, say $\mathcal{I}$, which could contain, for example, the desired common impact time. Different cooperative guidance commands correspond to different distributions over $\mathscr{G}$ for the adversary, which leads to different distributions over $\mathcal{O}$.
	
	Now, for the adversary, two state trajectories, $i_1 = \mathcal{P}_{\mathcal{O}\mid \mathbf{x}_1}$ and $i_2 = \mathcal{P}_{\mathcal{O}\mid \mathbf{x}_2}$ such that $i_1,i_2\in\mathcal{I}$ are $\varepsilon$-encrypted if
	\begin{align}
		\max_{i_1\neq i_2} \mathcal{T} \left(\mathcal{P}_{\mathcal{O}\mid \mathbf{x}_1},\mathcal{P}_{\mathcal{O}\mid \mathbf{x}_2}\right)\leq \varepsilon,
	\end{align}
	where $\mathcal{T}$ denotes the total-variation metric and $\mathcal{P}$ is the probability. This essentially means that small changes in $\mathbf{F}_k,\mathbf{B}_k$ produce proportionally small changes in the nominal observation, making it harder to distinguish, even though the vehicles are following different guidance principles.
	\begin{remark}
		Upon mid-engagement guidance morphing, the resultant momentary time-to-go errors among the pursuers reactivate the cooperative consensus protocol, which drives the pursuers to synchronize their time-to-go values again, ensuring that the new consensus value is reached within the prescribed time from the switching moment.
	\end{remark}
	\begin{corollary}
		Consider a cooperative target engagement scenario involving heterogeneous pursuers, wherein different pursuers exchange information over a directed graph $\mathcal{G}$, and adopt different guidance principles. The guidance command for the $k$\textsuperscript{th} pursuer guided by deviated pursuit guidance that leads to consensus in time-to-go within a predefined-time $t_e$, regardless of initial engagement configuration, is given by
		\begin{align}
			a_{P_k} =& {V_{P_k} \dot{{\theta}}_k} + \left[ \frac{V_{P_k} \left(V_{P_k}^2-{V}_{T}^2\right) \cos^2 {\delta}_k}{{r}_k {V}_{\theta_k}} \right]
			\left(\alpha - \beta \frac{\dot{h}(t)}{h(t)}\right) s_k\label{eq:amdp},
		\end{align}
		where $s_k=\displaystyle\sum _{j=1}^N[\mathcal{L}]_{kj} e_j$, for $k\in \{1,\ldots M\}$ and for some  $\alpha>0,\beta\geq{1}/({2\lambda_2(\mathcal{\hat{\mathcal{L}}})})$ while  $h(t) ,\dot{h}(t)$ are defined in \eqref{timevarying}-\eqref{timevaryingdot}. Consequently, the pursuers intercept the moving target simultaneously at a time $t_f$ decided via implicit coordination.
		
	\end{corollary}
	\begin{proof}
		For the pursuer guided by DPG, the time-to-go expression is given by \cite{hp4}
		\begin{equation}\label{eq:tgodp}
			t_{\mathrm{go}_k} = \dfrac{r_k\left(V_{r_k}+2V_{P_k}\cos\delta_k-V_{\theta_k}\tan\delta_k\right)}{V_{P_k}^2-V_T^2}.
		\end{equation}
		On differentiating \eqref{eq:tgodp} with respect to time we get,
		\begin{equation}
			\dot{t}_{\mathrm{go}_k} = -1 + \frac{{V}_{\theta_k}^2 \sec^2 {\delta}_k}{V_{P_k}^2-{V}_{T}^2} - \frac{{r}_k {V}_{\theta_k} \sec^2 {\delta}_k}{V_{P_k} \left(V_{P_k}^2-{V}_{T}^2\right)} a_{P_k}
			\label{eq:tgodynamicsdp}
		\end{equation}
		On comparing \eqref{eq:tgodynamicsdp} with \eqref{eq:tgo_affine} one may readily obtain
		\begin{align}\label{eq:FBdp}
			F_{k} = -1+\frac{V_{\theta_k}^2 \sec^2 \delta_k}{V_{P_k}^2 - V_{T}^2},\quad
			B_{k} = -\frac{r_k V_{\theta_k} \sec^2 \delta_k}{V_{P_k}\left(V_{P_k}^2 - V_{T}^2\right)}. 
		\end{align}
		By substituting \eqref{eq:FBdp} in \eqref{eq:aPk}, one may observe that the lateral acceleration $a_{P_k}$ conforms with \eqref{eq:amdp}. 
	\end{proof}
	\begin{remark}
		The term $s_k$ is composed of intra-class components $\displaystyle\sum _{j=1}^M[\mathcal{L}]_{kj} e_j$  and inter-class components  $\displaystyle\sum _{j=M+1}^N[\mathcal{L}]_{kj} e_j$. The first term corresponds to the contribution of pursuers within the same class, while the latter term represents the strategic injection of error information from dissimilar pursuer classes to ensure global coordination.
	\end{remark}
	\begin{remark}
		Once the consensus in time-to-go is established among pursuers, the term $s_k$ goes to zero, and from \eqref{eq:amdp}, one can obtain that $a_{P_k}=V_{P_k}\dot \theta_k$, which is a pursuit guidance term. Moreover, during the steady state, the condition $s_k=0$ drives $\dot{\delta}_i$ to zero, thereby maintaining the constant-bearing/deviation angle needed to anticipate the target's motion.
	\end{remark}
	\begin{corollary}
		Consider a cooperative target engagement scenario involving heterogeneous pursuers, wherein different pursuers exchange information over a directed graph $\mathcal{G}$, and adopt different guidance principles. The guidance command for the $k$\textsuperscript{th} pursuer guided by true proportional navigation guidance that leads to consensus in time-to-go within a predefined-time $t_e$, regardless of initial engagement configuration, is given by
		\begin{align}
			a_{M_k} &= c_k\,\dot{\theta}_k \notag\\       
			&\quad + \bigg[\frac{\bigl((V_{\theta_k})^2 + (V_{r_k})^2 + 2 c_k V_{r_k}\bigr)^2}
			{2\,(V_{r_k} + 2 c_k)\,V_{\theta_k}\,r_k}\bigg]
			\left(\alpha - \beta\,\frac{\dot{h}(t)}{h(t)}\right)\,s_k
			\label{eq:amtpn}
		\end{align}
		where $s_k=\displaystyle\sum _{j=1}^N[\mathcal{L}]_{kj} e_j$, for $k\in \{M+1,\ldots N\}$ and for some  $\alpha>0,\beta\geq{1}/({2\lambda_2(\mathcal{\hat{\mathcal{L}}})})$ while $h(t) ,\dot{h}(t)$ are defined in \eqref{timevarying}-\eqref{timevaryingdot}. Consequently, the pursuers intercept the moving target simultaneously at a time $t_f$ decided via implicit coordination.
	\end{corollary}
	\begin{proof}
		The time-to-go expression for a pursuer following TPNG is given by \cite{hp5}
		\begin{equation}\label{eq:tgotpn}
			t_{{\rm go}_k} = -\frac{r_k (V_{r_k} + 2c_k)}{V_{\theta_k}^2 + V_{r_k}^2 + 2c_k V_{r_k}},
		\end{equation}
		where $c_k$ is a design parameter that could be either constant or a function of the engagement parameters. On differentiating \eqref{eq:tgotpn} with respect to time, one may obtain
		\begin{align}
			\dot{t}_{{\rm go}_k}
			= & 
			-1 
			+ \frac{2 c_k V_{\theta_k}^2(V_{r_k} + 2 c_k) }{\bigl(V_{\theta_k}^2 + V_{r_k}^2 + 2 c_k V_{r_k}\bigr)^2} \notag \\
			& 
			- \frac{2 (V_{r_k} + 2 c_k) V_{\theta_k} r_k}{\bigl(V_{\theta_k}^2 + V_{r_k}^2 + 2 c_k V_{r_k}\bigr)^2} \, a_{P_k}.
			\label{eq:tgodynamicstpn}
		\end{align}
		On comparing \eqref{eq:tgodynamicstpn} with \eqref{eq:tgo_affine} it follows that
		\begin{align}
			F_{k} &= -1 
			+ \frac{2 c_k V_{\theta_k}^2(V_{r_k} + 2 c_k) }{\bigl(V_{\theta_k}^2 + V_{r_k}^2 + 2 c_k V_{r_k}\bigr)^2}, \label{eq:Ftpn}\\
			B_{k} &= - \frac{2 (V_{r_k} + 2 c_k) V_{\theta_k} r_k}{\bigl(V_{\theta_k}^2 + V_{r_k}^2 + 2 c_k V_{r_k}\bigr)^2} .\label{eq:Btpn}
		\end{align}
		On substituting \eqref{eq:Ftpn}-\eqref{eq:Btpn} in \eqref{eq:aPk}, the cooperative guidance command $a_{P_k}$ in \eqref{eq:amtpn} could be obtained.
	\end{proof}

	\section{Simulations}
	
\begin{figure}[htbp]
		\centering
		\begin{subfigure}{0.42\linewidth}
			\centering
			\includegraphics[width=\linewidth]{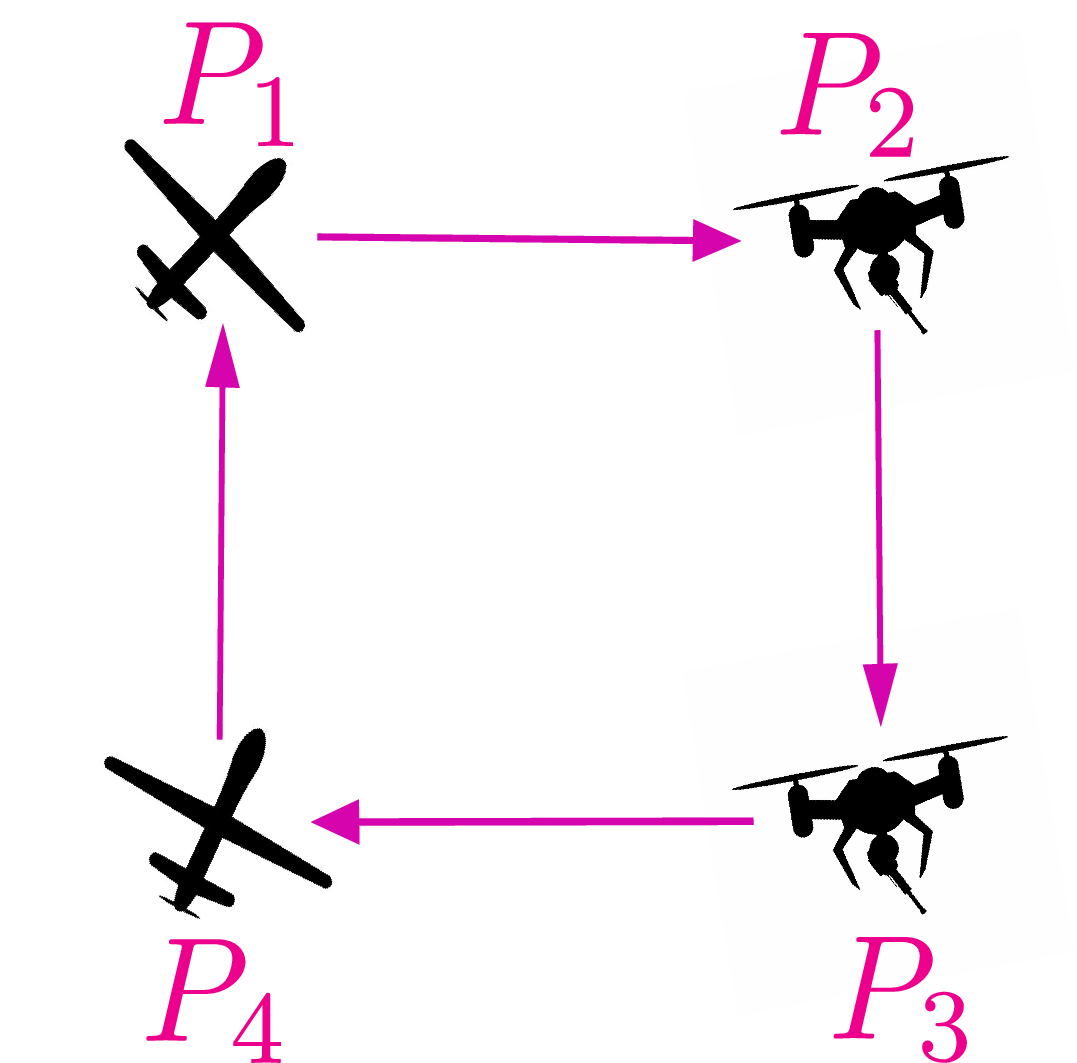}
			\caption{Topology 1.}
			\label{fig:topology_a}
		\end{subfigure}
		\hfill
		\begin{subfigure}{0.42\linewidth}
			\centering
			\includegraphics[width=\linewidth]{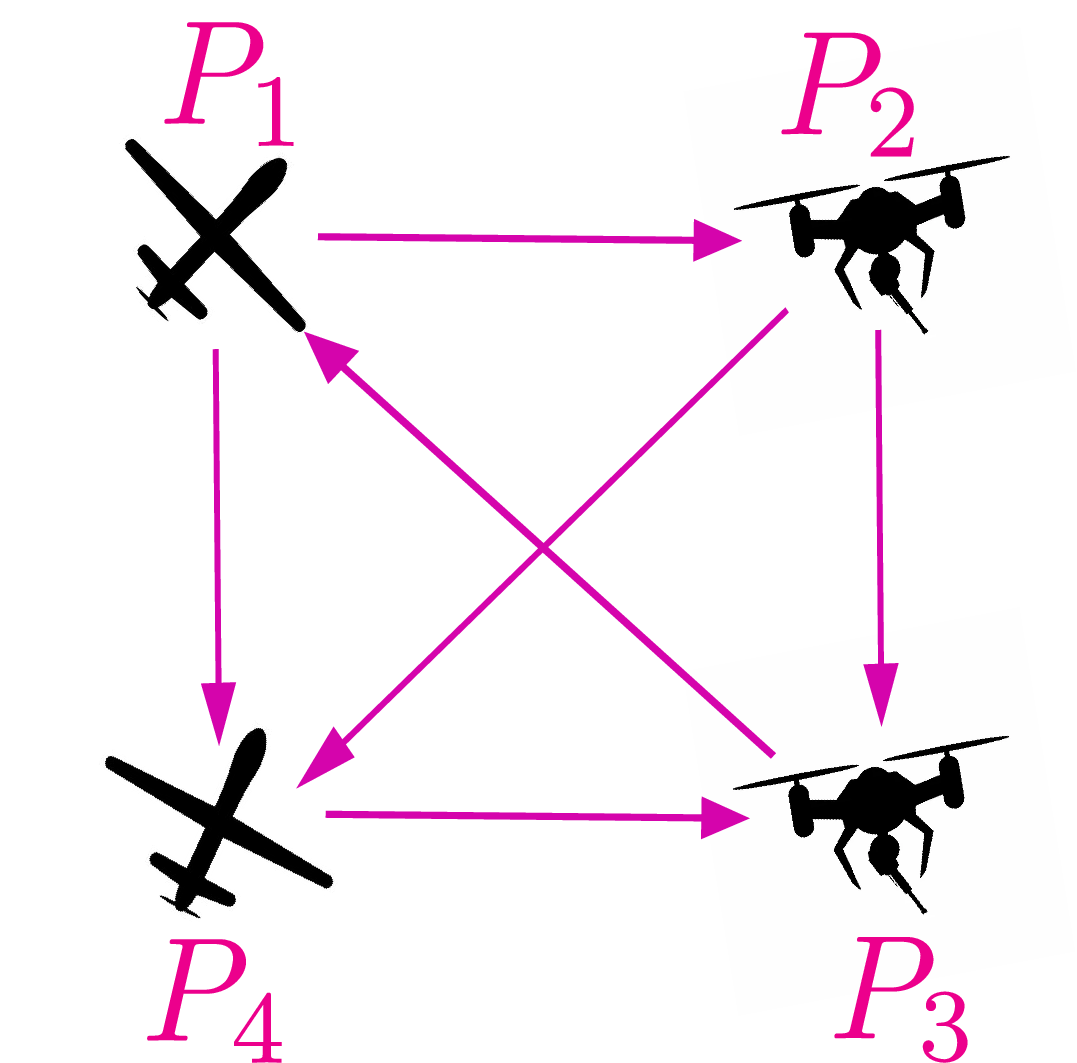}
			\caption{Topology 2.}
			\label{fig:topology_b}
		\end{subfigure}
		\caption{Communication Topologies}
		\label{fig:topologies}
	\end{figure}

	We consider a cooperative engagement scenario involving a team of four pursuers, labeled $P_1$ through $P_4$. This constitutes a heterogeneous system where the agents employ different guidance strategies. Pursuers $P_1$ and $P_4$ utilize DPG, while $P_2$ and $P_3$ adopt TPNG. The team operates under fixed, directed communication topologies, as illustrated in \Cref{fig:topologies}. It is assumed that the $V_{P_{\max}}>V_T$.  To establish a heterogeneous swarm, all four pursuers are launched from the origin with distinct initial speeds and heading angles. The speed vector is given by $\mathbf{V}_{p} = \begin{bmatrix} 73 & 55 & 54 & 70 \end{bmatrix}^\top \,\text{m/s}$, and the corresponding heading angle vector is $\boldsymbol{\gamma}_{p} = \begin{bmatrix} 10^\circ & 5^\circ & 18^\circ & 12^\circ \end{bmatrix}^\top$, thereby ensuring heterogeneity in both speed and direction of motion. The target is launched from the position $(1000,\,0)$ m with a heading angle of $60^\circ$ and a speed of $55$ m/s. The parameter $c_k$ is considered as $c_k=3(V_{P_k}+V_T)$. The initial time-to-go estimates of the pursuers (based on their respective strategies) are computed as  $\mathbf{t}_{{\rm go}}(0) = \begin{bmatrix} 40.463 & 41.000 & 44.768 & 47.434 \end{bmatrix}^\top$ s. 
	
	In the first case, pursuers communicate over Topology 1 (a directed cycle) as illustrated in  \Cref{fig:topology_a}. In accordance with the design requirements, the controller gains are chosen as $\alpha = 2$ and $\beta = 1.2$, while the prescribed convergence time is set as $t_e = 5\,\text{s}$.The pursuers are subject to actuator constraints, with their maximum allowable acceleration limited to $6g $, where $g$ denotes the acceleration due to gravity.
	\begin{figure*}[h!]
		\centering
		\begin{subfigure}[t]{0.475\linewidth}
			\centering
			\includegraphics[width=\linewidth]{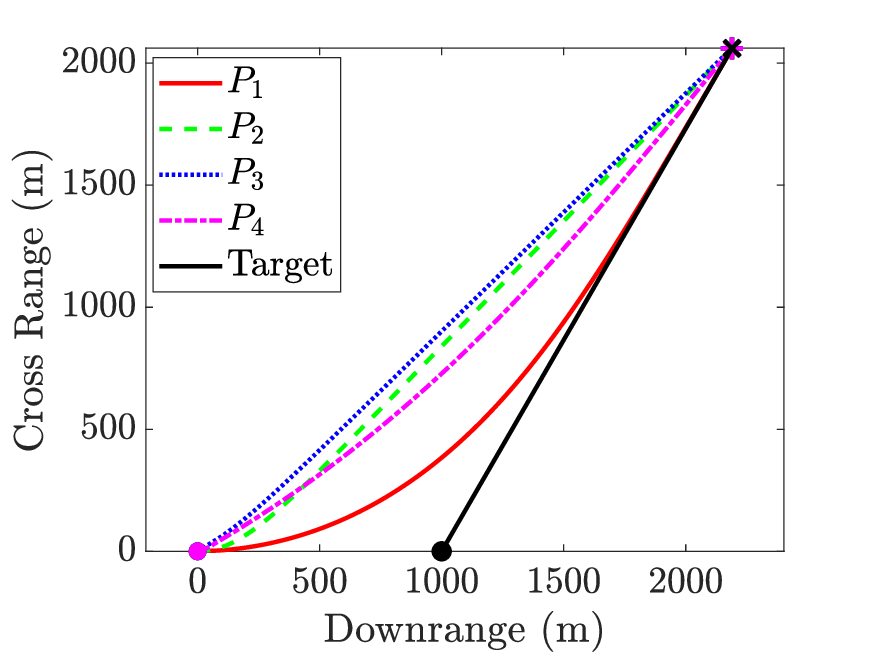}
			\caption{Trajectories.}
			\label{fig:traj1}
		\end{subfigure}
		\begin{subfigure}[t]{0.475\linewidth}
			\centering
			\includegraphics[width=\linewidth]{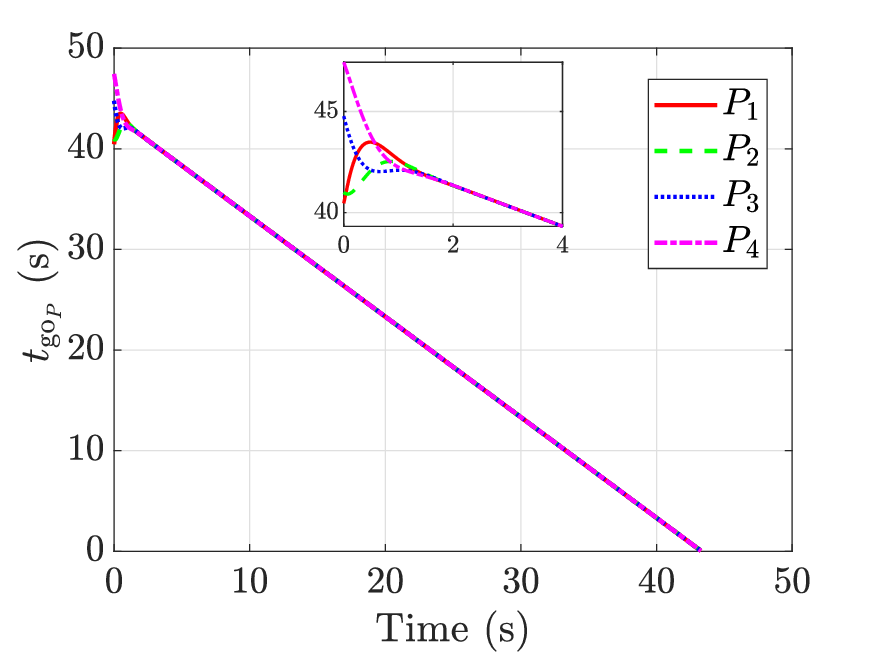}
			\caption{Time-to-go.}
			\label{fig:tgo1}
		\end{subfigure}
		\begin{subfigure}[t]{0.475\linewidth}
			\centering
			\includegraphics[width=\linewidth]{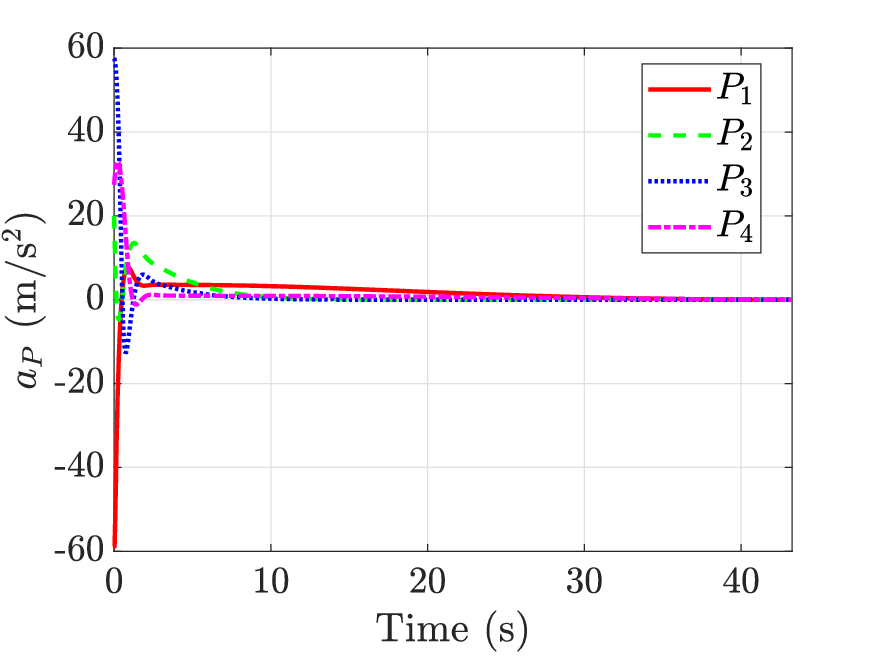}
			\caption{Accelerations.}
			\label{fig:am1}
		\end{subfigure}
		\begin{subfigure}[t]{0.475\linewidth}
			\centering
			\includegraphics[width=\linewidth]{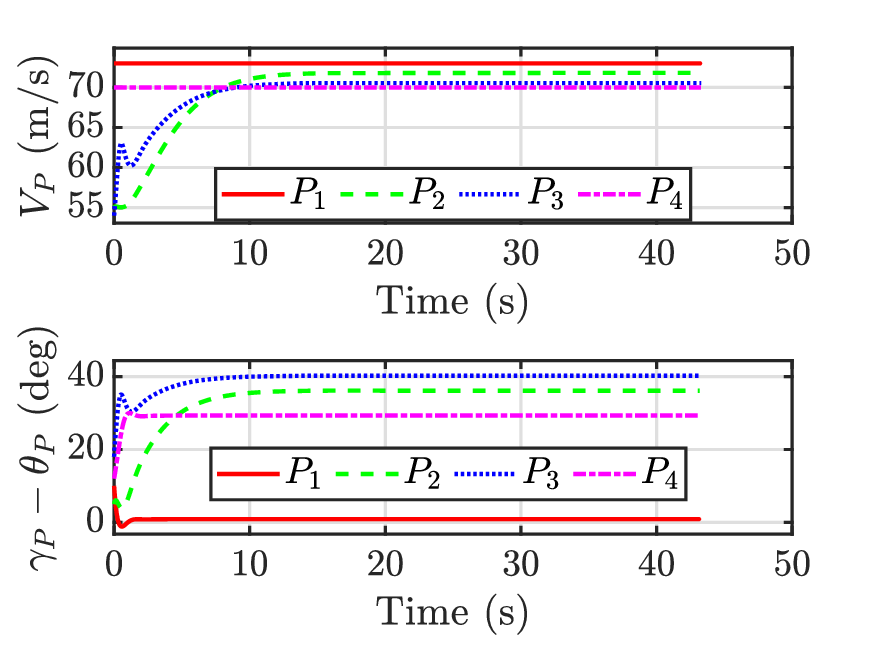}
			\caption{Velocity and Lead Angle Variation.}
			\label{fig:vh1}
		\end{subfigure}
		\caption{Performance of the proposed method for a typical case.}
		\label{fig:performance_1}
	\end{figure*}
	
	\Cref{fig:traj1}  illustrates the diversity in pursuers' trajectories arising from the heterogeneity of guidance principles, which leads to a coordinated interception of the target at approximately 43.3 s.  The DPG pursuers exhibit highly curved trajectories, whereas the TPNG pursuers follow relatively straighter paths. This inherent diversity acts as an effective encryption of the trajectories, making them unpredictable to the adversary. One may observe from \Cref{fig:tgo1} that, regardless of heterogeneity, all pursuers attained consensus in time-to-go within 3 s, which is significantly less than the user-specified upper bound $t_e$. From \Cref{fig:am1} we can observe that initially acceleration demand is higher until consensus is achieved, after which the magnitude continuously decreases as the engagement proceeds, eventually leading to zero terminal demand, which is a highly beneficial characteristic. One can observe from \Cref{fig:vh1},  that once consensus is achieved, the deviation angle stabilizes for $P_1$ and $P_4$. For $P_2$ and $P_3$, the lead angle remains constant as the radial acceleration component decreases to zero, corresponding to a constant velocity.

\begin{figure*}[h!]
		\centering
		\begin{subfigure}[t]{0.475\linewidth}
			\centering
			\includegraphics[width=\linewidth]{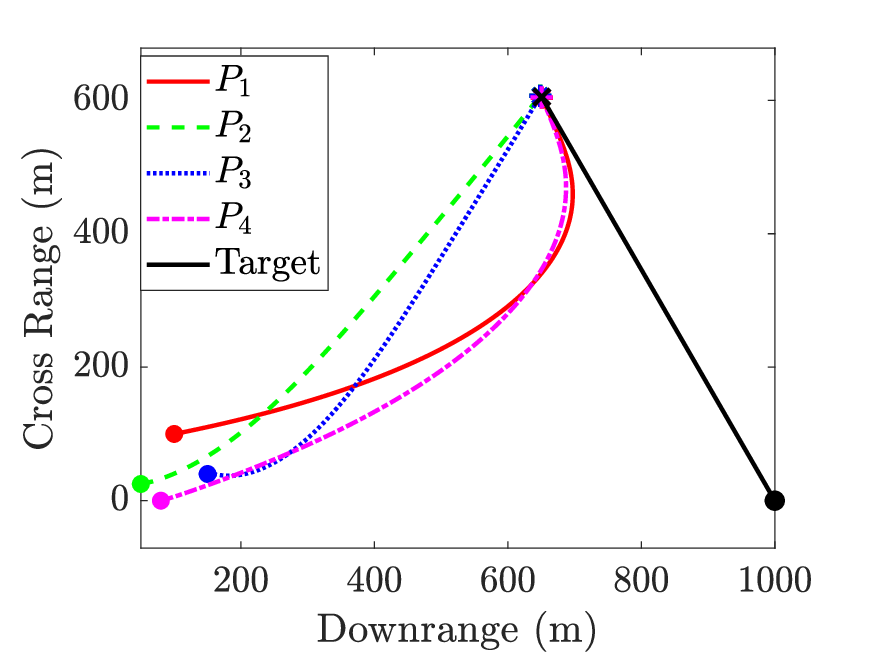}
			\caption{Trajectories.}
			\label{fig:traj2}
		\end{subfigure}
		\begin{subfigure}[t]{0.475\linewidth}
			\centering
			\includegraphics[width=\linewidth]{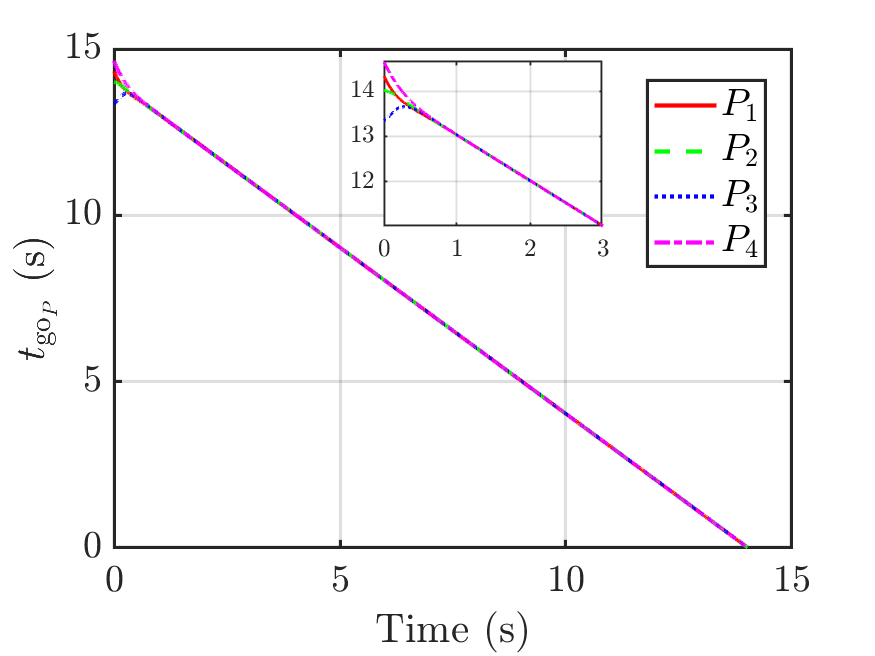}
			\caption{Time-to-go.}
			\label{fig:tgo2}
		\end{subfigure}
		\begin{subfigure}[t]{0.475\linewidth}
			\centering
			\includegraphics[width=\linewidth]{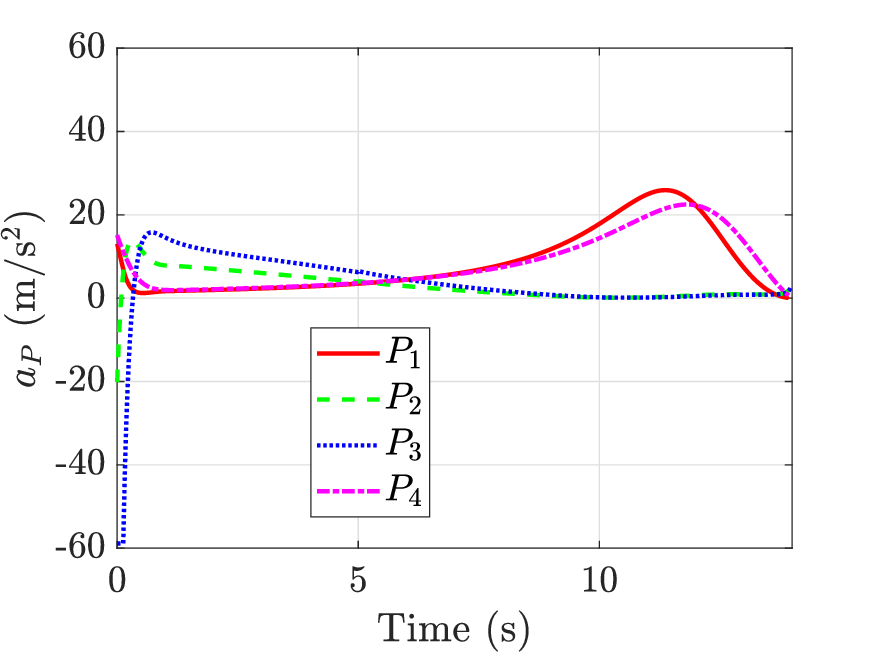}
			\caption{Accelerations.}
			\label{fig:am2}
		\end{subfigure}
		\begin{subfigure}[t]{0.475\linewidth}
			\centering
			\includegraphics[width=\linewidth]{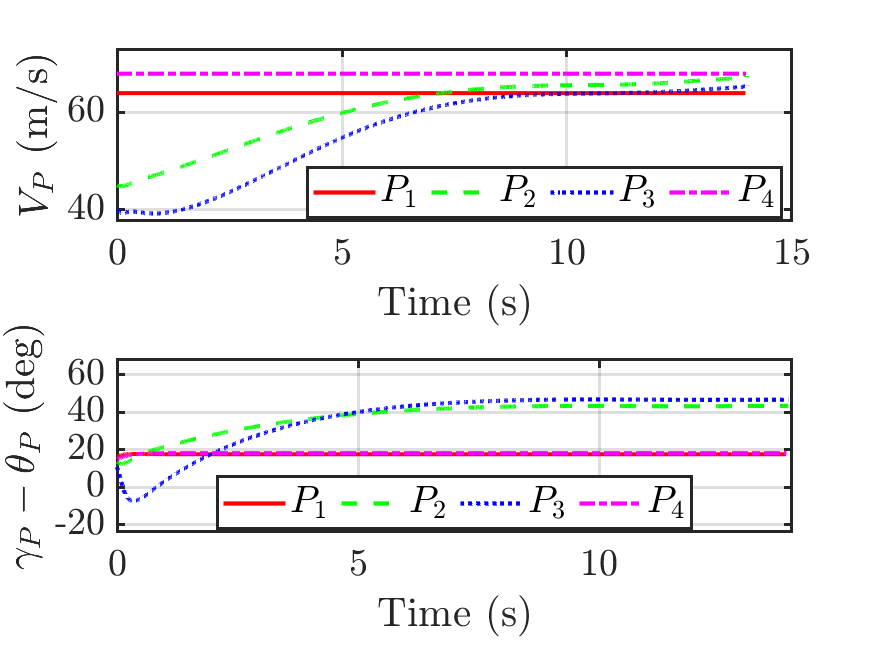}
			\caption{Velocity and Lead Angle Variation.}
			\label{fig:vh2}
		\end{subfigure}
		\caption{Efficacy of the proposed method for new engagement geometry.}
		\label{fig:performance_2}
	\end{figure*}

	The proposed strategy's performance is further evaluated in an alternative scenario characterized by a different engagement geometry and communication graph (Topology 2 in \Cref{fig:topology_b}). In this case, the pursuers are launched from different initial positions, each with a unique speed and heading angle. Compared to the earlier cycle graph, where the four pursuers ($P_1$--$P_4$) relied on a simple sequential structure in which information could only propagate along a single directed loop, the new graph \Cref{fig:topology_b} exhibits a densely connected topology. The presence of multiple paths and redundant communication links (for example, both a direct $P_1 \to P_4$ link and an indirect $P_1 \to P_2 \to P_4$ path) reduces potential information delays, enhances resilience against link failures, and enables faster consensus and synchronization. This structural advantage substantially improves the reliability and effectiveness of cooperative salvo engagements. In this case, the target is initiated from the position $(1000,\,0)$ m with a heading angle of $120^\circ$ and a velocity of $50$ m/s. The pursuers are deployed from the initial positions  
	$\begin{bmatrix} 
		(100,\,100) &
		(50,\,25) &
		(150,\,40) &
		(80,\,0) 
	\end{bmatrix}\,\text{m}$. 
	The corresponding speeds and heading angles of the pursuers are defined as 
	$\mathbf{V}_{p} = 
	\begin{bmatrix} 
		64 & 45 & 40 & 68 
	\end{bmatrix}^\top 
	\, \text{m/s}, \quad 
	\boldsymbol{\gamma}_{p} = 
	\begin{bmatrix} 
		10^\circ & 12^\circ & 8^\circ & 15^\circ 
	\end{bmatrix}^\top$. The initial time-to-go estimates are given by  $\mathbf{t}_{{\rm go}}(0) = \begin{bmatrix} 14.330& 14.012 & 13.342 & 14.639 \end{bmatrix}^\top \, \text{s}$
	The controller parameters and actuator constraints are assumed to be identical to those used in the previous scenario. One may observe from \Cref{fig:traj2} that despite the changes in the engagement geometry, diversity in the trajectories followed by the pursuers is evident, and all the pursuers cooperatively intercepted the target at approximately  14 s. Similar to the previous engagement, consensus in time-to-go is achieved well within the upper bound $t_e$. The abrupt increase in lateral acceleration demand for pursuers $P_1$ and $P_4$ arises from the sharp maneuver induced by the pursuit component of the DPG guidance command, after which the demand gradually decreases to zero in the terminal phase.
	
	\begin{figure*}[h!]
		\centering
		\begin{subfigure}[t]{0.475\linewidth}
			\centering
			\includegraphics[width=\linewidth]{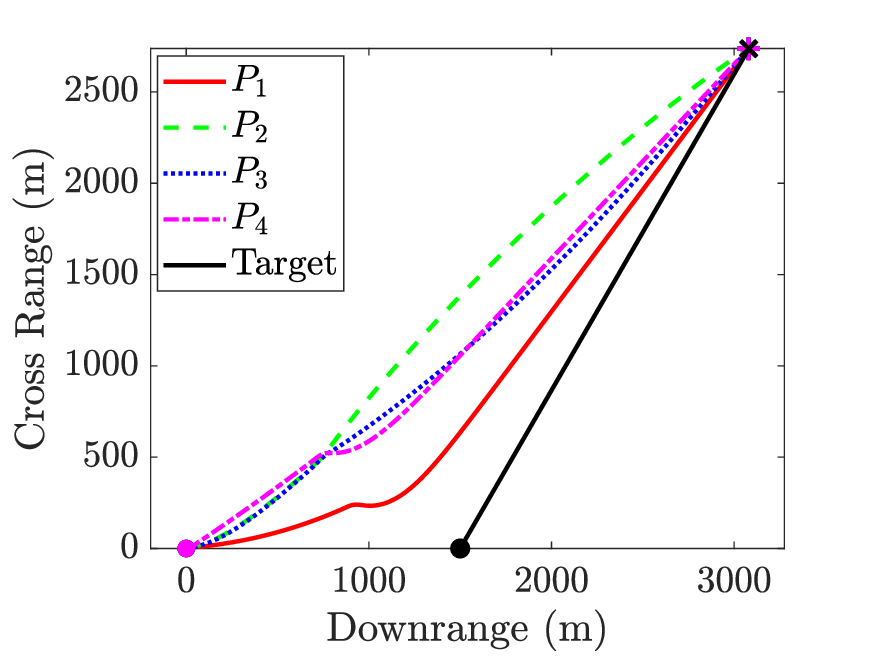}
			\caption{Trajectories.}
			\label{fig:traj3}
		\end{subfigure}
		\begin{subfigure}[t]{0.475\linewidth}
			\centering
			\includegraphics[width=\linewidth]{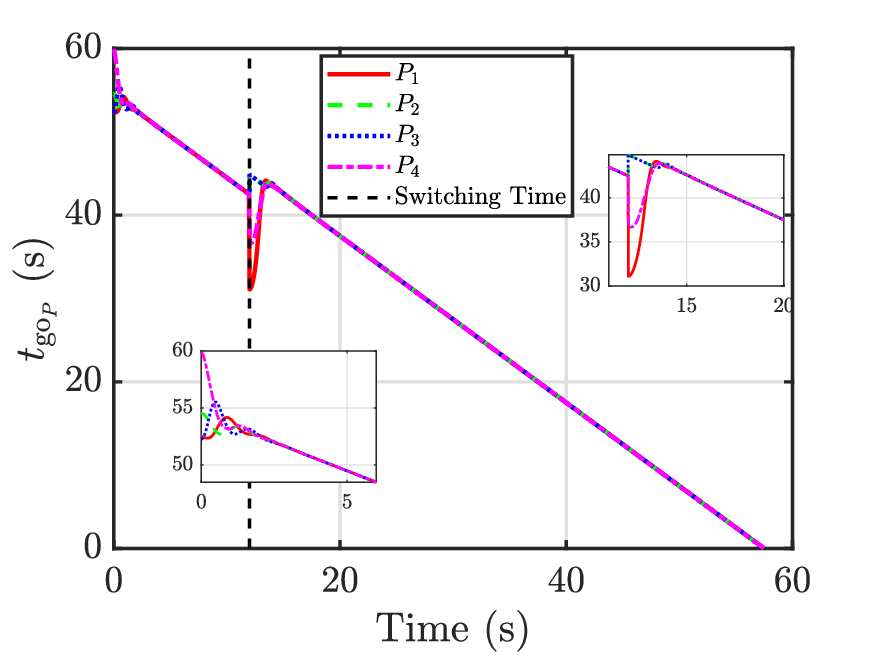}
			\caption{Time-to-go.}
			\label{fig:tgo3}
		\end{subfigure}
		\begin{subfigure}[t]{0.475\linewidth}
			\centering
			\includegraphics[width=\linewidth]{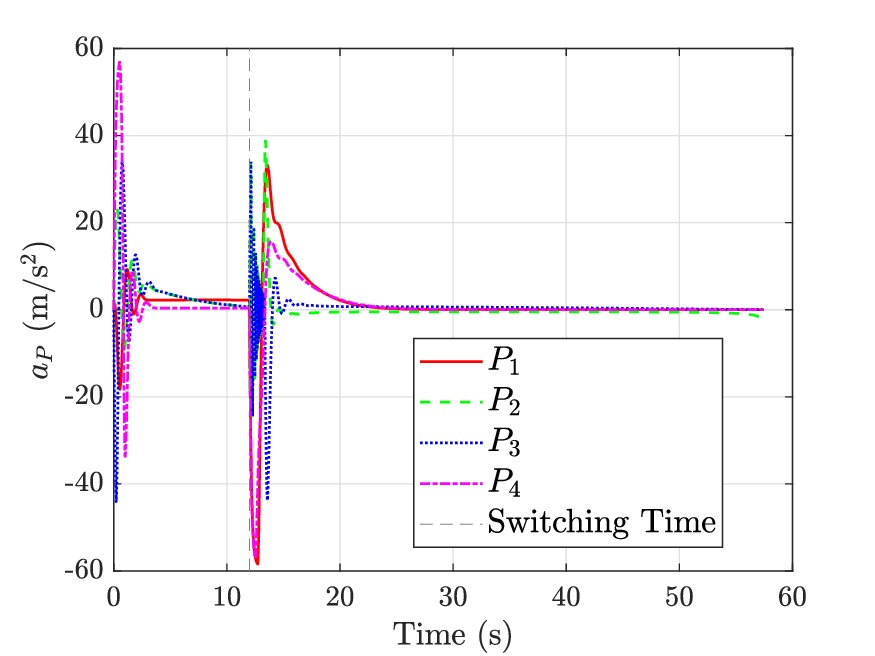}
			\caption{Accelerations.}
			\label{fig:am3}
		\end{subfigure}
		\begin{subfigure}[t]{0.475\linewidth}
			\centering
			\includegraphics[width=\linewidth]{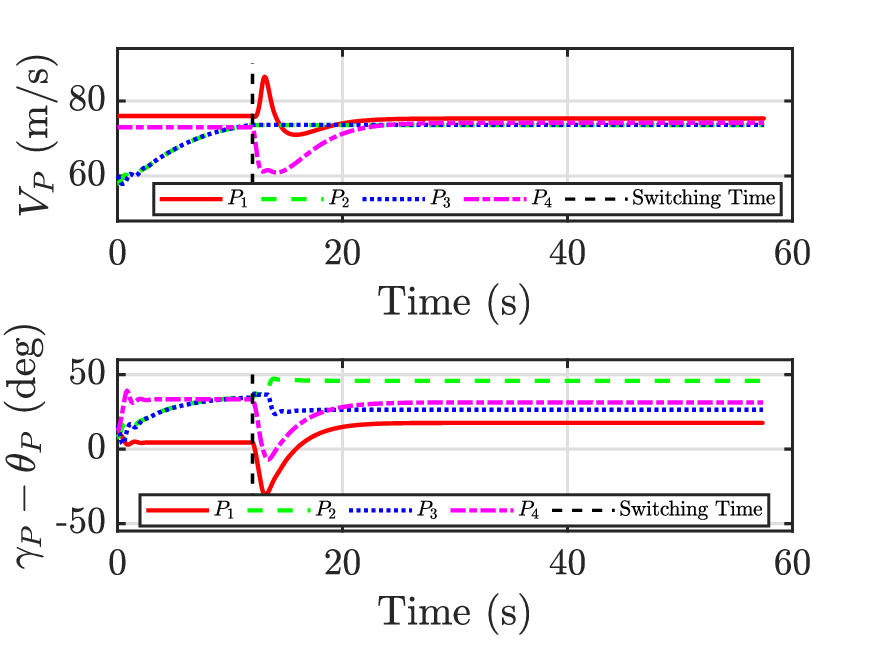}
			\caption{Velocity and Lead Angle Variation.}
			\label{fig:vh3}
		\end{subfigure}
		\caption{Efficacy of the proposed method under guidance morphing.}
		\label{fig:performance_3}
	\end{figure*}
	To demonstrate the efficacy of the switching capability, we assume that all vehicles are multirotors with the ability to switch between TPNG and DPG. Since multirotors can generate both lateral and radial accelerations, they are well-suited for such transitions. The pursuers are launched from the origin with speeds and heading angles specified as 
	$\mathbf{V}_{p} = 
	\begin{bmatrix} 
		76 & 58 & 60 & 73 
	\end{bmatrix}^\top 
	\, \text{m/s}, \quad 
	\boldsymbol{\gamma}_{p} = 
	\begin{bmatrix} 
		10^\circ & 8^\circ & 16^\circ & 12^\circ 
	\end{bmatrix}^\top$. 
	During the initial phase of engagement, pursuers $P_1$ and $P_4$ employ DPG, while pursuers $P_2$ and $P_3$ follow TPNG. At the switching instant $t_{\text{switch}} = 12 \; \text{s}$, the guidance modes are interchanged such that $P_1$ and $P_4$ transition to TPNG, whereas $P_2$ and $P_3$ switch to DPG. The target is launched from the position $(1500,0)\,\text{m}$ with a heading angle $\gamma_T = 60^\circ$ and a speed of $V_T = 55\,\text{m/s}$. The controller gains $\alpha$ and $\beta$ are both set as 2 and prescribed time $t_e$ is taken as $10\;\text{s}$. The parameter $c_k$ and the actuator constraints are assumed to be identical to the previous scenario. The initial time-to-go estimates are given by  $\mathbf{t}_{{\rm go}}(0) = \begin{bmatrix} 52.494 & 54.504 & 52.246 & 59.900 \end{bmatrix}^\top \, \text{s}$. 
	
	As shown in \Cref{fig:traj3}, all pursuers achieve cooperative interception of the target at $57.5 \, \text{s}$ despite guidance morphing. By allowing the same pursuer to follow multiple guidance laws over time, the resulting trajectory diversity exceeds that of a heterogeneous group alone, adding an inherent layer of trajectory encryption that makes prediction by an adversary significantly more difficult. It can be observed from \Cref{fig:tgo3} that the pursuers initially reach consensus in their time-to-go values at approximately $4 \, \text{s}$, well within the prescribed time $t_e$. Following the guidance morphing, momentary disagreement arises in the time-to-go values, prompting the reactivation of the cooperative consensus protocol, which drives the pursuers’ time-to-go values to synchronize once again within $5 \text{s}$. The commanded accelerations are filtered through a first-order low-pass filter with a time constant of $0.15 \, \text{s}$ and unity DC gain, and the resulting accelerations are shown in \Cref{fig:am3}. The acceleration demand is initially high but gradually decreases as the pursuers achieve consensus. Following guidance morphing, it rises again to accommodate the new guidance requirements and the reactivation of the cooperative consensus protocol, ultimately gradually decreasing to zero during the terminal phase. From \Cref{fig:vh3} it can be observed that the velocities and lead angles of the pursuers exhibit behavior similar to the previous scenarios, consistent with the guidance law being followed.
	\begin{figure*}[h!]
		\centering
		\begin{subfigure}[t]{0.475\linewidth}
			\centering
			\includegraphics[width=\linewidth]{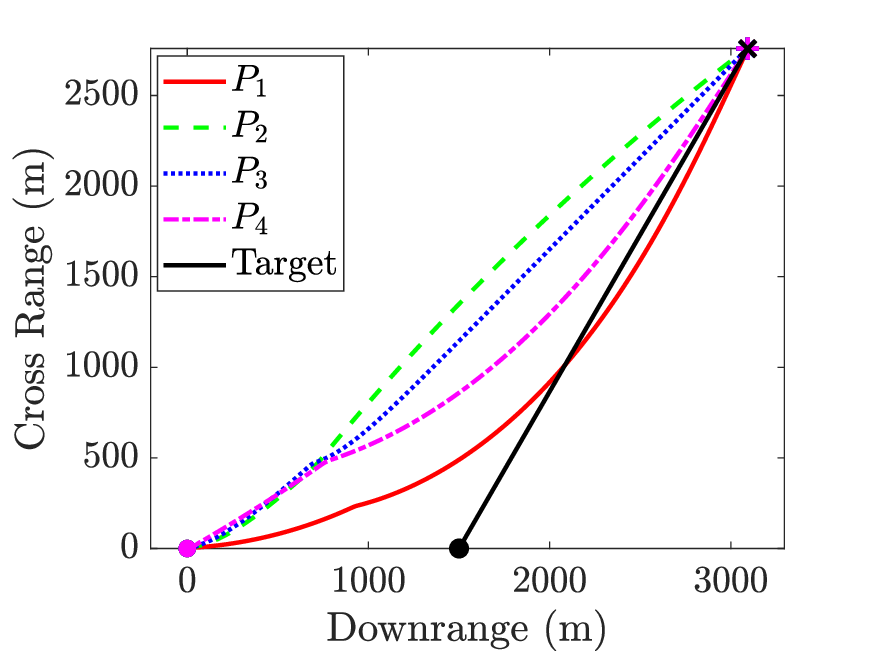}
			\caption{Trajectories.}
			\label{fig:traj4}
		\end{subfigure}
		\begin{subfigure}[t]{0.475\linewidth}
			\centering
			\includegraphics[width=\linewidth]{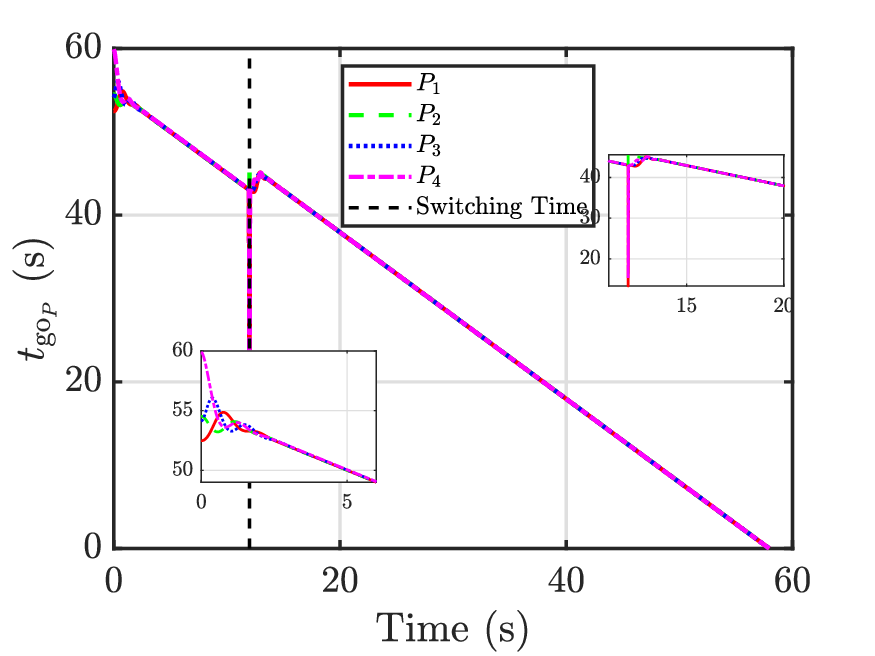}
			\caption{Time-to-go.}
			\label{fig:tgo4}
		\end{subfigure}
		\begin{subfigure}[t]{0.475\linewidth}
			\centering
			\includegraphics[width=\linewidth]{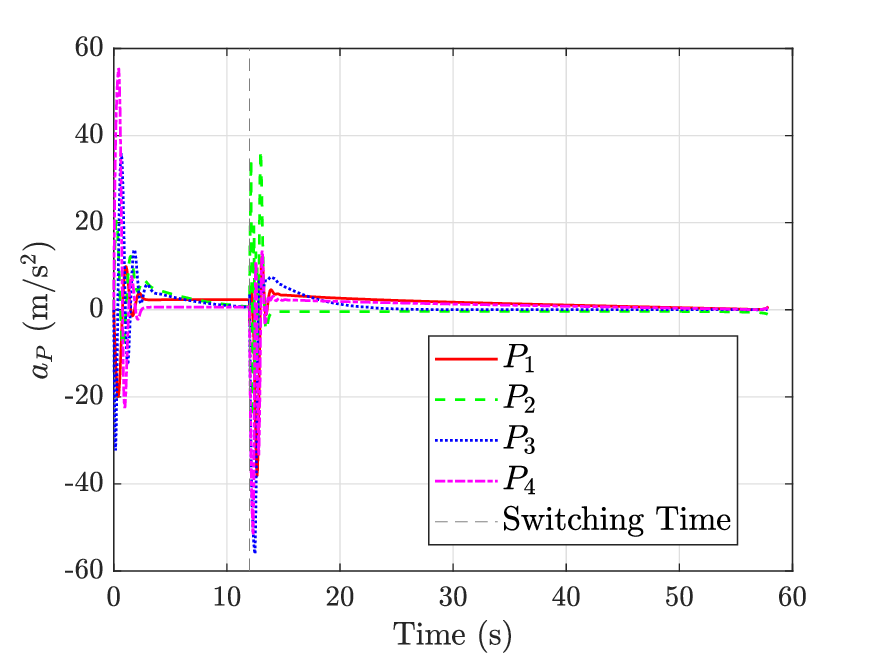}
			\caption{Accelerations.}
			\label{fig:am4}
		\end{subfigure}
		\begin{subfigure}[t]{0.475\linewidth}
			\centering
			\includegraphics[width=\linewidth]{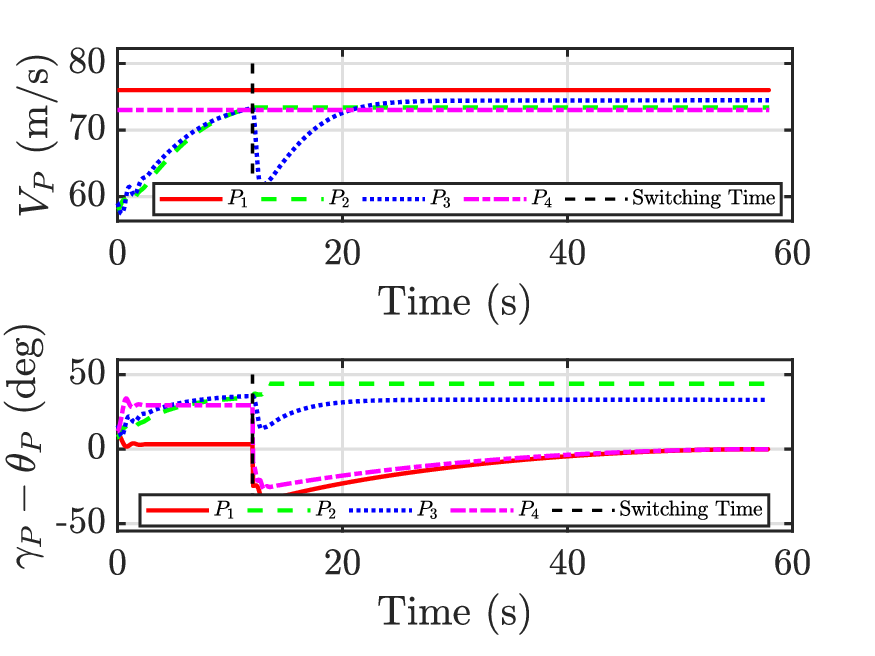}
			\caption{Velocity and Lead Angle Variation.}
			\label{fig:vh4}
		\end{subfigure}
		\caption{Efficacy of the proposed method under guidance morphing with PIP, DPG, and TPNG.}
		\label{fig:performance_4}
	\end{figure*}
	
	The proposed strategy is further validated without assuming the vehicle is a multirotor, we employ the Predicted Interception Point (PIP)–based guidance strategy proposed in \cite{sp1} (requires control solely through lateral acceleration channel), in conjunction with DPG and TPNG. Consider the same engagement conditions as in the previous case, but pursuers $P_1$ and $P_4$ switch to the PIP-based guidance strategy, $P_2$ employs DPG, and $P_3$ does not switch and follows TPNG. From \Cref{fig:traj4}, it can be observed that the pursuers intercept the target cooperatively at $58 \text{s}$. Compared to the previous case, this demonstrates that cooperative interception remains feasible even when three different guidance laws are followed. Similar behaviors in the time-to-go and acceleration demands can be observed from \Cref{fig:tgo4} and \Cref{fig:am4}, respectively, consistent with the previous case.  Pursuers $P_2$ and $P_3$ exhibit velocity and lead angle characteristics consistent with the guidance law being followed, as shown in \Cref{fig:vh4}. In contrast, for $P_1$ and $P_4$, the velocity remains constant after switching, while the lead angle ultimately converges to zero, which is consistent with \cite{sp1}.

	\section{Conclusions and Future Work}\label{sec:conclusions}
	In this paper, we presented a cooperative guidance framework for simultaneous target interception that leverages heterogeneity in guidance principles as a strategic design feature. The effectiveness of the strategy is demonstrated primarily for DPG and TPNG guidance laws, showing that heterogeneous time-to-go formulations enable cooperative simultaneous interception. Furthermore, additional simulations with the PIP guidance strategy illustrate that multiple vehicles can follow different guidance laws simultaneously while still achieving cooperative interception. The swarm generates diverse trajectory families through heterogeneous guidance strategies, enabling flexible adjustment of impact timing while concealing the collective interception intent as a trajectory-domain encryption layer, which facilitates coordination under different guidance laws and increases the unpredictability of swarm dynamics from an adversarial perspective. Future work will extend this approach to maneuvering targets, time-varying communication topologies, and stochastic disturbances.

	\bibliographystyle{IEEEtran}
	\bibliography{referencesnew}
	
\end{document}